\pdfoutput=1
\documentclass[prd,aps,twocolumn,superscriptaddress,nofootinbib,amsmath,amssymb,10pt]{revtex4-2}
\usepackage[T1]{fontenc}
\usepackage{lmodern}
\usepackage{bm}
\usepackage{microtype}
\usepackage{tikz}
\usepackage[hidelinks]{hyperref}
\DeclareMathOperator{\supp}{supp}

\begin{document}
\title{Induced Bell--CHSH observables in the Fewster--Verch measurement scheme}
\author{M.~S.~Guimaraes}
\email{msguimaraes@uerj.br}
\affiliation{UERJ -- Universidade do Estado do Rio de Janeiro, Instituto de F\'isica -- Departamento de F\'isica Te\'orica, Rua S\~ao Francisco Xavier 524, 20550-013, Maracan\~a, Rio de Janeiro, Brazil}
\author{I.~Roditi}
\email{roditi@cbpf.br}
\affiliation{CBPF -- Centro Brasileiro de Pesquisas F\'isicas, Rua Dr.~Xavier Sigaud 150, 22290-180, Rio de Janeiro, Brazil}
\author{S.~P.~Sorella}
\email{silvio.sorella@fis.uerj.br}
\affiliation{UERJ -- Universidade do Estado do Rio de Janeiro, Instituto de F\'isica -- Departamento de F\'isica Te\'orica, Rua S\~ao Francisco Xavier 524, 20550-013, Maracan\~a, Rio de Janeiro, Brazil}
\begin{abstract}
The Summers--Werner theorem shows that the vacuum of a relativistic quantum field theory violates the Bell--CHSH inequality maximally on complementary wedges, but it is an existence statement: it exhibits neither the observables nor an apparatus that measures them. We give an explicit operator-level realization within the Fewster--Verch measurement framework, obtaining dichotomic Bell observables of a free Majorana field in $1+1$ dimensions as observables induced by local couplings to Majorana probes. The induced system smearing factorizes exactly as $f_h=k\,b_{\lambda,h}$, so its support is controlled by the compactly supported coupling profile $k$; dichotomy is imposed by calibrating and normalizing the induced smearings rather than required of the probe observable; and a probe-only normal ordering yields the product of Alice's and Bob's induced bilinears, which the non-multiplicative induced map alone does not provide. Under explicit attainability hypotheses the induced observables saturate the Tsirelson bound in a state prepared from the vacuum. The construction extends to coupled scalar fields.
\end{abstract}
\maketitle
\section{Introduction}
\label{sec:intro}
Measurement in relativistic quantum field theory requires an account of both the observable being measured and the spacetime process through which information about it is obtained. A historical motivation for examining this distinction is Sorkin's 1993 analysis of ``impossible measurements'': a straightforward extension of ideal state-reduction prescriptions to quantum fields can permit superluminal signalling \cite{Sorkin1993}. This difficulty motivates a description in which the interaction with the measuring apparatus, its localization, and the subsequent readout are specified explicitly.

Fewster and Verch developed an algebraic framework in which a measurement is described by a local interaction between the quantum field system and a prepared quantum field probe \cite{FV2020}. Its central contribution is to relate the observable read out on the probe to an induced observable of the system, so that the measurement is specified through the coupling, probe preparation and readout. The framework also describes the conditional change of the system state and relates the composition of measurement operations to their causal arrangement. This formulation is compatible with a generally covariant treatment on curved spacetimes, as explained in \cite{Fewster2019}, and provides the basis for the construction used here. Subsequent work has clarified its scope: the superluminal signalling of Sorkin's examples cannot arise from any measurement scheme with a locally coupled probe \cite{BFR2021}; the state updates it induces are causal for a broad class of local operations \cite{Jubb2022}; every local observable admits an asymptotic measurement scheme \cite{FJR2023}; and its relation to the particle-detector models widely used in relativistic quantum information has been analysed in \cite{PoloGomez2022}.

In the Fewster--Verch (FV) scheme, the system and probe are quantum field theories coupled in a compact spacetime region. The resulting scattering map, together with the initial probe state, associates outgoing probe observables with induced system observables through a completely positive map. The general framework establishes localization in the causal hull of the coupling region and describes conditional state changes using measurement instruments. Under causal factorization, causally ordered measurements compose consistently; spacelike separated coupling regions allow either ordering \cite{FV2020,Fewster2019}. These properties provide an operational setting in which to ask how locally coupled probes can generate the observables entering a Bell--CHSH construction.

Since the work of Summers and Werner \cite{SummersWerner1985,SummersWerner1987}, maximal violation of the Bell--CHSH inequality in quantum field theory has been established at the level of general principles: the vacuum restricted to the algebras of complementary wedges violates the inequality maximally, but the argument is an existence statement resting on the modular structure of the local algebras and does not exhibit the observables that achieve it. The same is true of the general FV realizability results for local POVMs \cite{Fewster2026,Mandrysch2026}. The gap between such structural statements and an explicit realization is considerable: it is only at the operator level that one can ask which smearings, which couplings and which probe readouts produce a given Bell setting, and it is at this level that the present construction is situated. For the free massive Majorana field in $1+1$ dimensions, explicit even bilinears of smeared fields localized in the right and left wedges have been shown to bring the vacuum Bell--CHSH correlator arbitrarily close to Tsirelson's bound, with test functions generated from compact modular seeds through the Tomita--Takesaki operator of the wedge \cite{MajoranaBilinears2026}. The present article shows how dichotomic Majorana bilinears of this type can be defined as output observables of local FV probe couplings.

The present article addresses this question for a Majorana system coupled locally to Majorana probes. Its main results are the following.
\begin{enumerate}
\item[(i)] \emph{Exact localization.} The system smearing induced by a probe observable has the factorized form
\begin{equation}
f_h(x)=k(x)b_{\lambda,h}(x),
\label{eq:intro-factorization}
\end{equation}
where \(k\in C_0^\infty\) is the interaction profile. Hence \(\operatorname{supp}f_h\subseteq\operatorname{supp}k\): the induced observable admits a smearing representative supported entirely within the coupling region (Sec.~\ref{sec:model}, Appendix~\ref{app:moller}).
\item[(ii)] \emph{Induced dichotomic bilinears.} Probe bilinears induce Hermitian Majorana bilinears of the system. Dichotomy is not required of the probe observable; it is obtained for the induced observable by calibrating the induced smearings to be real-orthogonal and normalizing them in the CAR norm (Sec.~\ref{sec:construction}).
\item[(iii)] \emph{Joint statistics and non-multiplicativity.} The induced-observable map is completely positive but not multiplicative, so first-moment identities cannot be multiplied to obtain joint Alice--Bob statistics. A probe-only normal ordering removes the probe contractions from the four-field expression and yields exactly the product of the two induced bilinears (Sec.~\ref{sec:normalordering}).
\item[(iv)] \emph{Conditional Tsirelson saturation.} Under explicit attainability hypotheses the induced bilinears generate two commuting local Pauli algebras, and in a state prepared from the vacuum by a nonzero joint projection the CHSH expectation equals \(2\sqrt2\) (Sec.~\ref{sec:bell}, Appendix~\ref{app:pauli}).
\item[(v)] \emph{Bosonic extension.} The construction extends to coupled massive scalar fields through sign--cosine Weyl readouts (Appendix~\ref{app:bosonic}).
\end{enumerate}

The article is organized as follows. Section~\ref{sec:model} introduces the Majorana system and probe and their local interaction. Section~\ref{sec:construction} develops the construction of induced dichotomic Majorana bilinears, including the calibration and normalization of the induced observables. Section~\ref{sec:normalordering} uses probe-only contraction subtraction to construct quartic system expressions and prepare the passage to Bell--CHSH correlations. Section~\ref{sec:bell} assembles the induced Bell--CHSH structure and, in Sec.~\ref{sec:maximal}, summarizes the conditional maximal CHSH expectation. Section~\ref{sec:discussion} synthesizes the construction, its conditional maximal-correlation result and its operational limitations, and identifies further questions about attainable ranges and probe readout. Appendix~\ref{app:moller} presents the exact factorization obtained from the iterative M\o ller solution. Appendix~\ref{app:pauli} constructs the local Pauli algebras, the singlet state from the vacuum and the saturation of the Tsirelson bound. Appendix~\ref{app:bosonic} extends the analysis to coupled massive scalar fields using the odd-harmonic sign--cosine construction of \cite{FiniteWeyl2026}. It derives the probe-vacuum cancellation for finite normally ordered Weyl sums and the sign--cosine limit of their induced system observables, then constructs Pauli algebras and a conditional singlet under attainable symplectic-pair assumptions. The existence of a bounded readout for the complete normally ordered probe series is kept separate from this induced limit.

\section{Majorana system and probe}
\label{sec:model}
Let \(\psi\) and \(\chi\) denote independent Majorana system and probe fields. For definiteness, take two massive free Majorana fields in two-dimensional Minkowski spacetime. We work with their graded field algebras, so odd system and probe fields anticommute. Physical bilinears belong to the even subalgebras. For real Majorana smearings, adopt

\begin{equation}
\begin{gathered}
\psi(f)^*=\psi(f),\qquad
\{\psi(f),\psi(f')\}=2r_\psi(f,f')\mathbf1,\\
r_\psi(f,f')=\operatorname{Re}\langle f,f'\rangle_\psi,
\end{gathered}
\end{equation}
\begin{equation}
\begin{gathered}
\chi(g)^*=\chi(g),\qquad
\{\chi(g),\chi(g')\}=2r_\chi(g,g')\mathbf1,\\
\{\psi(f),\chi(g)\}=0.
\end{gathered}
\end{equation}

The norm \(\|f\|_\psi^2=r_\psi(f,f)\) is the CAR norm of the physical smearing class, after quotienting null smearings; it is not an arbitrary spacetime \(L^2\) norm. The same convention applies to the probe. We suppress the algebra embeddings when writing fields in the combined algebra.

We fix the spinor conventions in $1+1$ dimensions, with $x^0=t$, $x^1=x$, $\hbar=c=1$, and $\eta^{\mu\nu}=\operatorname{diag}(1,-1)$. A Majorana representation is \cite{DeVincenzo2020}
\begin{equation}
\begin{gathered}
\gamma^0=\sigma_2=\begin{pmatrix}0&-i\\i&0\end{pmatrix},\qquad
\gamma^1=-i\sigma_3=\begin{pmatrix}-i&0\\0&i\end{pmatrix},\\
\{\gamma^\mu,\gamma^\nu\}=2\eta^{\mu\nu}\mathbf1_2.
\end{gathered}
\label{eq:gamma}
\end{equation}
Here $(\gamma^0)^\dagger=\gamma^0$, $(\gamma^1)^\dagger=-\gamma^1$, and both $i\gamma^\mu$ are real. Define the Dirac conjugate and fix the charge-conjugation phase by
\begin{equation}
\begin{gathered}
\bar\zeta=\zeta^\dagger\gamma^0,\qquad
C=-\gamma^0,\qquad C^T=-C,\\
C(\gamma^\mu)^TC^{-1}=-\gamma^\mu,\qquad
\zeta^c=C\bar\zeta^{\,T}=\zeta^*.
\end{gathered}
\label{eq:charge}
\end{equation}
Indeed, $C(\gamma^0)^T=\mathbf1_2$, so the Majorana conditions $\psi^c=\psi$ and $\chi^c=\chi$ mean real classical Grassmann components, or componentwise self-adjoint quantum fields. The distinction between the matrix $C$ and the antilinear charge-conjugation operation is essential \cite{Stone2020}. For a Majorana field, $\bar\zeta=\zeta^T\gamma^0=-\zeta^TC$. Proper boosts are represented by the real matrices $L(\theta)=\exp(\theta\gamma^0\gamma^1/2)=\exp(\theta\sigma_1/2)$, with $L^T\gamma^0L=\gamma^0$; hence $\bar\psi\chi$ is a Lorentz scalar.

Take positive masses $m_\psi,m_\chi$, without requiring equality. With real $\lambda$ and real $k\in C_0^\infty(O)$, the complete action is
\begin{equation}
\begin{aligned}
S&=S_0+S_{\mathrm{int}},\\
S_0&=\frac12\int d^2x\,\big[
\bar\psi(i\gamma^\mu\partial_\mu-m_\psi)\psi\\
&\hphantom{=\frac12\int d^2x\,\big[}
+\bar\chi(i\gamma^\mu\partial_\mu-m_\chi)\chi\big],\\
S_{\mathrm{int}}&=\lambda\int d^2x\,k(x)\bar\psi(x)\chi(x)\\
&=\frac\lambda2\int d^2x\,k(x)(\bar\psi\chi+\bar\chi\psi).
\end{aligned}
\label{eq:action}
\end{equation}
The factor $1/2$ in each free action is the Majorana factor. The interaction has no additional factor $1/2$ after combining its two equal cross terms. For anticommuting species and $(\gamma^0)^T=-\gamma^0$, one has $\bar\chi\psi=\bar\psi\chi$ and $(\bar\psi\chi)^\dagger=\bar\chi\psi$; the interaction is therefore Hermitian. The free action is Hermitian after integration by parts, with vanishing boundary terms. Equivalently, its kinetic term may be written with $i\overleftrightarrow\partial_\mu/2$ inside the brackets.

The real contraction matrix in a component expression with a separate factor $i$ is now fixed, rather than arbitrary:
\begin{equation}
\begin{gathered}
M=-i\gamma^0=iC=\begin{pmatrix}0&-1\\1&0\end{pmatrix},\\
S_{\mathrm{int}}=i\lambda\int d^2x\,k\psi^TM\chi.
\end{gathered}
\label{eq:M}
\end{equation}
Thus $M\ne C$: $C$ is imaginary in this convention, whereas $M$ is real. Explicitly, $\bar\psi\chi=i(-\psi_1\chi_2+\psi_2\chi_1)$, and $(i\psi_a\chi_b)^\dagger=i\psi_a\chi_b$. For $\Xi=(\psi,\chi)^T$, the same interaction is $(i\lambda/2)\int d^2x\,\Xi^T\bigl(\begin{smallmatrix}0&kM\\-kM^T&0\end{smallmatrix}\bigr)\Xi$.

Put $D_j=i\gamma^\mu\partial_\mu-m_j$ for $j=\psi,\chi$. Variation with odd variations placed on the left gives, after integration by parts,
\begin{equation}
\begin{aligned}
\delta S=\int d^2x\,\big[&
\delta\psi^T\gamma^0(D_\psi\psi+\lambda k\chi)\\
&+\delta\chi^T\gamma^0(D_\chi\chi+\lambda k\psi)\big],
\end{aligned}
\end{equation}
so that
\begin{equation}
D_\psi\psi+\lambda k\chi=0,\qquad
D_\chi\chi+\lambda k\psi=0.
\label{eq:eom}
\end{equation}
The two equal contributions from varying a free Majorana quadratic form cancel its factor $1/2$; moving the variation of the cross term to the left uses both the Grassmann sign and the antisymmetry of $\gamma^0$. In species space the scalar mass matrix is $\bigl(\begin{smallmatrix}m_\psi&-\lambda k\\-\lambda k&m_\chi\end{smallmatrix}\bigr)$. Since $D_j$ and this mixing are real, the coupled equations preserve the Majorana reality condition. Real retarded and advanced Green operators consequently preserve real spinors, and their real dual maps preserve the component smearings used below. We use $\psi(f)=\int d^2x\,f^T\psi$ for real component test spinors, and similarly for $\chi$; the CAR norms already specified are norms on their physical classes, not spacetime $L^2$ norms.

Assume that the resulting coupled dynamics has a well-defined, parity-preserving scattering isomorphism \(\Theta\), with the convention that it pulls outgoing observables back to the incoming uncoupled algebra. Quadratic dynamics gives linear field transformations. For an allowed real outgoing probe smearing \(h\), write

\begin{equation}
\begin{aligned}
X_h:=\Theta(\mathbf1\widehat\otimes\chi(h))
&=\chi_{\mathrm{in}}(G_\lambda h)+\psi_{\mathrm{in}}(F_\lambda h)\\
&=\chi_{\mathrm{in}}(g_h)+\psi_{\mathrm{in}}(f_h).
\end{aligned}
\end{equation}

The hat denotes the graded tensor product. The maps \(F_\lambda,G_\lambda\) are real-linear on the allowed Majorana smearings and preserve the adjoint convention. Because \(\Theta\) preserves the CAR,

\begin{equation}
r_\chi(h,h')
=r_\chi(g_h,g_{h'})+r_\psi(f_h,f_{h'}).
\label{eq:carid}
\end{equation}

Thus the two parts of the output field are constrained by the same scattering transformation. In this model the factorization discussed in Appendix~\ref{app:moller} is written

\begin{equation}
f_h(x)=k(x)b_{\lambda,h}(x),\qquad
\operatorname{supp}f_h\subseteq\operatorname{supp}k.
\end{equation}

This identifies a representative of the induced smearing with controlled support; it does not by itself prove that every smearing with that support is in the range of \(F_\lambda\).

Prepare the incoming system and probe independently in a graded product state \(\omega\widehat\otimes\sigma\), where \(\sigma\) is even. In particular, every odd probe field has zero expectation. Taking the probe expectation defines \(\eta_\sigma\); on system-first ordered monomials its action is

\begin{equation}
\begin{aligned}
&\eta_\sigma\big(\psi(f_1)\cdots\psi(f_m)
\chi(g_1)\cdots\chi(g_n)\big)\\
&\qquad=\psi(f_1)\cdots\psi(f_m)\,
\sigma\big(\chi(g_1)\cdots\chi(g_n)\big).
\end{aligned}
\end{equation}

For an even probe observable \(B\), the induced system observable is

\begin{equation}
\begin{gathered}
\varepsilon_\sigma(B)
=\eta_\sigma\big(\Theta(\mathbf1\widehat\otimes B)\big),\\
(\omega\widehat\otimes\sigma)\big(\Theta(\mathbf1\widehat\otimes B)\big)
=\omega\big(\varepsilon_\sigma(B)\big).
\end{gathered}
\end{equation}

The scattering map is multiplicative, whereas the induced-observable map is generally not. Products must therefore be expanded before taking the probe expectation \cite{FV2020}.

\subsection{Alice--Bob system}
\label{sec:alice-bob}
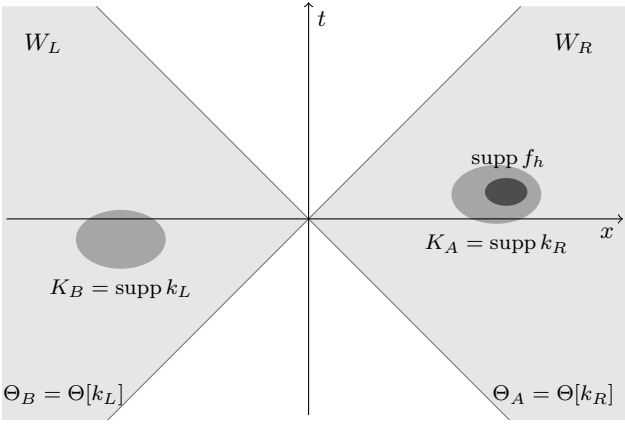
\begin{figure}[t]
\centering
\small
\begin{tikzpicture}[scale=1.08]
\clip (-3.75,-2.45) rectangle (3.95,2.75);
\fill[black!10] (0,0) -- (2.6,2.6) -- (3.9,2.6) -- (3.9,-2.6) -- (2.6,-2.6) -- cycle;
\fill[black!10] (0,0) -- (-2.6,2.6) -- (-3.9,2.6) -- (-3.9,-2.6) -- (-2.6,-2.6) -- cycle;
\draw[black!55] (-2.6,-2.6) -- (2.6,2.6);
\draw[black!55] (-2.6,2.6) -- (2.6,-2.6);
\fill[black!35] (2.3,0.3) ellipse (0.55 and 0.36);
\fill[black!35] (-2.3,-0.25) ellipse (0.55 and 0.36);
\fill[black!70] (2.42,0.33) ellipse (0.26 and 0.17);
\node[below,font=\footnotesize] at (2.3,-0.08) {$K_A=\supp k_R$};
\node[below,font=\footnotesize] at (-2.3,-0.63) {$K_B=\supp k_L$};
\node[above,font=\footnotesize] at (2.45,0.52) {$\supp f_h$};
\draw[->] (-3.7,0) -- (3.85,0) node[below left] {$x$};
\draw[->] (0,-2.4) -- (0,2.65) node[below right] {$t$};
\node at (3.25,2.15) {$W_R$};
\node at (-3.25,2.15) {$W_L$};
\node[font=\footnotesize] at (3.0,-2.15) {$\Theta_A=\Theta[k_R]$};
\node[font=\footnotesize] at (-3.0,-2.15) {$\Theta_B=\Theta[k_L]$};
\end{tikzpicture}
\caption{Spacetime arrangement of the Alice--Bob configuration. The coupling profiles $k_R$ and $k_L$ have compact supports $K_A\subset W_R$ and $K_B\subset W_L$ (shaded ellipses) inside the spacelike separated right and left wedges. By the exact factorization $f_h=k\,b_{\lambda,h}$ the observable induced by a probe observable read out after Alice's coupling admits a smearing representative supported entirely within $K_A$ (dark region). Since $K_A$ and $K_B$ are causally disjoint, the scattering automorphisms commute, $\Theta=\Theta_A\circ\Theta_B=\Theta_B\circ\Theta_A$, and each acts trivially on observables localized in the other wedge.}
\label{fig:wedges}
\end{figure}

For the Bell configuration, assign Alice to the right wedge and Bob to the left wedge of the two-dimensional Minkowski spacetime used above:
\begin{equation}
W_R=\{(t,x):x>|t|\},\qquad
W_L=\{(t,x):x<-|t|\}.
\end{equation}
These wedges are spacelike separated. In the interaction of Eq.~\eqref{eq:action}, choose
\begin{equation}
\begin{gathered}
k(x)=k_R(x)+k_L(x),\\
k_R\in C_0^\infty(W_R),\qquad k_L\in C_0^\infty(W_L),
\end{gathered}
\label{eq:wedge-couplings}
\end{equation}
with real profiles. Thus $K_A=\supp k_R$ and $K_B=\supp k_L$ are compact coupling regions inside the respective wedges, rather than interactions extending throughout either wedge. The geometry is shown in Fig.~\ref{fig:wedges}. The interaction splits as
\begin{equation}
\begin{gathered}
S_{\mathrm{int}}=S_{\mathrm{int},A}+S_{\mathrm{int},B},\\
S_{\mathrm{int},A}=\lambda\int d^2x\,k_R\bar\psi\chi,\qquad
S_{\mathrm{int},B}=\lambda\int d^2x\,k_L\bar\psi\chi.
\end{gathered}
\end{equation}

This displays the geometric split for the single probe field of Eq.~\eqref{eq:action}. When independent Alice and Bob probes are used for the joint readout below, the same local interaction is applied to two independent probe copies $\chi_A,\chi_B$, with $\chi$ replaced by $\chi_A$ in the first term and by $\chi_B$ in the second. Their free actions and graded probe algebras are then included separately; independence of the two probes is an additional preparation choice, not a consequence of the wedge geometry.

The interaction densities are even: each contains a product of two odd Majorana fields. Graded locality therefore makes the Alice and Bob densities commute at spacelike separation. With the causal factorization property of the FV scattering construction, the scattering automorphism for the combined coupling factors into the automorphisms for the two local couplings \cite{FV2020,Fewster2019}:
\begin{equation}
\begin{gathered}
\Theta[k_R+k_L]=\Theta_A\circ\Theta_B
=\Theta_B\circ\Theta_A,\\
\Theta_A=\Theta[k_R],\qquad\Theta_B=\Theta[k_L].
\end{gathered}
\label{eq:alice-bob-factorization}
\end{equation}
For independent probes, each factor in Eq.~\eqref{eq:alice-bob-factorization} is extended by the identity on the other probe algebra. All factors use the convention already fixed here: outgoing observables are pulled back to the incoming uncoupled algebra. Their composition is the algebraic description of the interaction evolution; no Hilbert-space implementer is assumed. Since $K_A$ and $K_B$ are causally disjoint, either causal ordering is admissible and the factors commute.

Locality also specifies precisely what each factor leaves unchanged. If $K^\perp$ denotes the spacetime region causally disjoint from $K$, the Alice scattering automorphism acts as the identity on observables localized in $K_A^\perp$, and similarly for Bob. In particular, for local observables $O_B$ and $O_A$ supported in $W_L$ and $W_R$, respectively,
\begin{equation}
\Theta_A(O_B)=O_B,\qquad \Theta_B(O_A)=O_A.
\end{equation}
Thus Alice's coupling does not change Bob's spacelike separated local observables, and conversely. The corresponding non-selective local measurement operation leaves remote system expectations unchanged. Conditional post-selection can nevertheless change those expectations through pre-existing correlations \cite{FV2020}; causal independence of the couplings neither removes Alice--Bob correlations nor requires a product system state. The locality statement concerns regions causally disjoint from the coupling support, not arbitrary later observables in its causal future. Together with the attainable-range assumptions imposed below, this geometry supplies the separated local observables entering the Bell--CHSH construction.

\section{Construction of induced dichotomic Majorana bilinears}
\label{sec:construction}

\textbf{The starting probe observable is not required to be an involution.} Here ``dichotomic'' means a normalized Hermitian involution with spectral values $\pm1$. The corrected probe bilinear may also have two spectral values, but these are generally different from $\pm1$. Subtraction of the probe contraction, calibration and output normalization are included in the construction below.

\textbf{Standing assumptions.} The following hypotheses are used throughout and are referred to by number. (A1)~The M\o ller map of the coupled theory is well defined and the iterative solution of Appendix~\ref{app:moller} can be resummed, so that $\Theta$ is an automorphism of the graded incoming algebra and $f_h=kb_{\lambda,h}$ holds exactly rather than order by order. (A2)~The probe preparation $\sigma$ is an even state, so that odd probe expectations vanish; for the joint readout the two probe copies are independently prepared. (A3)~\emph{Attainability:} the induced smearings $f_h$ entering a given construction are non-null in the CAR norm and, when two or three are needed on one side, linearly independent. (A4)~In the Alice--Bob configuration the coupling regions satisfy $K_A\subset W_R$, $K_B\subset W_L$, so that they are causally disjoint. (A1) is a hypothesis on the model and (A3) a hypothesis on the range of the local output map, discussed in Sec.~\ref{sec:calibration} and Sec.~\ref{sec:discussion}; (A2) and (A4) are choices of the protocol.

\subsection{Expansion of the outgoing product}

Consider two probe smearings \(h,h'\) and the corresponding outgoing fields, written in the incoming algebra:
\begin{equation}
\chi_{\mathrm{out}}(h)=\chi_{\mathrm{in}}(g_h)+\psi_{\mathrm{in}}(f_h),
\end{equation}
\begin{equation}
\chi_{\mathrm{out}}(h')=\chi_{\mathrm{in}}(g_{h'})+\psi_{\mathrm{in}}(f_{h'}).
\end{equation}
These are the two decompositions used throughout the calculation below.

Choose real allowed probe smearings \(h,h'\), and abbreviate

\begin{equation}
\begin{gathered}
p=\chi_{\mathrm{in}}(g_h),\qquad p'=\chi_{\mathrm{in}}(g_{h'}),\\
s=\psi_{\mathrm{in}}(f_h),\qquad s'=\psi_{\mathrm{in}}(f_{h'}).
\end{gathered}
\end{equation}

The output decomposition gives \(X_h=p+s\) and \(X_{h'}=p'+s'\). Keeping the original order first,

\begin{equation}
X_hX_{h'}=pp'+ps'+sp'+ss'.
\end{equation}

Moving each mixed monomial into system-first order uses the cross-species CAR once:

\begin{equation}
ps'=-s'p,\qquad
X_hX_{h'}=ss'+pp'-s'p+sp'.
\end{equation}

Since the probe state is even,

\begin{equation}
\sigma(p)=\sigma(p')=0,
\qquad c_\sigma(h,h'):=\sigma(pp'),
\end{equation}

and hence

\begin{equation}
\begin{gathered}
\eta_\sigma(ss')=ss',\qquad
\eta_\sigma(pp')=c_\sigma(h,h')\mathbf1,\\
\eta_\sigma(-s'p)=\eta_\sigma(sp')=0.
\end{gathered}
\end{equation}

It follows that

\begin{equation}
\boxed{\varepsilon_\sigma\big(\chi(h)\chi(h')\big)
=\psi(f_h)\psi(f_{h'})+c_\sigma(h,h')\mathbf1.}
\end{equation}

The mixed terms vanish because of the probe one-point functions, independently of the system one-point functions. The two-point contraction generally remains, including in the probe vacuum. Quasifreeness is not required for this two-field calculation.

The raw product \(\chi(h)\chi(h')\) is an even algebra element but need not be Hermitian. It is a starting point for the calculation, not yet the proposed measurement observable. The induced-product formula also explicitly demonstrates the failure of multiplicativity:

\begin{equation}
\begin{gathered}
\varepsilon_\sigma(\chi(h))\varepsilon_\sigma(\chi(h'))=ss',\\
\varepsilon_\sigma(\chi(h)\chi(h'))=ss'+c_\sigma\mathbf1.
\end{gathered}
\end{equation}

These identities for odd fields are understood in the graded field extension; the eventual measurement observable below is even.

\subsection{Probe contraction subtraction and Hermiticity}

Define the subtraction on the expanded outgoing expression by

\begin{equation}
\mathcal N_\sigma[X_hX_{h'}]
:=X_hX_{h'}-\sigma\big(\chi_{\mathrm{in}}(g_h)
\chi_{\mathrm{in}}(g_{h'})\big)\mathbf1.
\end{equation}

Only the incoming-probe contraction is subtracted. Thus

\begin{equation}
\eta_\sigma\big(\mathcal N_\sigma[X_hX_{h'}]\big)=ss'.
\end{equation}

The same expression is the scattering image of the calibrated probe-algebra element

\begin{equation}
\begin{gathered}
Q_\sigma(h,h')=\chi(h)\chi(h')-c_\sigma(h,h')\mathbf1,\\
\varepsilon_\sigma(Q_\sigma)=ss'.
\end{gathered}
\end{equation}

This subtraction depends on both the prepared probe state and the scattering map through \(g_h,g_{h'}\). It must not be silently replaced by ordinary normal ordering of \(\chi(h)\chi(h')\) using \(\sigma(\chi(h)\chi(h'))\); the two contractions are generally different.

To display a Hermitian observable before imposing orthogonality, put

\begin{equation}
r_f=r_\psi(f_h,f_{h'}),\qquad r_g=r_\chi(g_h,g_{h'}).
\end{equation}

The probe CAR and the adjoint property of a state imply

\begin{equation}
\begin{gathered}
c_\sigma(h,h')+\overline{c_\sigma(h,h')}=2r_g,\\
c_\sigma(h,h')=r_g+i d_\sigma,\qquad d_\sigma\in\mathbb R.
\end{gathered}
\end{equation}

Consider the Hermitian probe bilinear

\begin{equation}
\begin{aligned}
B_0(h,h')&=\frac{i}{2}[\chi(h),\chi(h')]\\
&=i\big(\chi(h)\chi(h')-r_\chi(h,h')\mathbf1\big).
\end{aligned}
\end{equation}

Using the scattering CAR identity and the induced-product formula, rather than a multiplicativity assumption, gives

\begin{equation}
\begin{gathered}
\varepsilon_\sigma(B_0)
=i(ss'-r_f\mathbf1)+\beta_\sigma\mathbf1,\\
\beta_\sigma=i(c_\sigma-r_g)=-d_\sigma\in\mathbb R.
\end{gathered}
\end{equation}

Subtract this real probe offset:

\begin{equation}
\begin{gathered}
B_{\mathrm{corr}}=B_0-\beta_\sigma\mathbf1,\\
\varepsilon_\sigma(B_{\mathrm{corr}})
=i(ss'-r_f\mathbf1)=\frac{i}{2}[s,s'].
\end{gathered}
\end{equation}

Both sides are Hermitian. If \(r_f=0\), then the scattering CAR identity gives \(r_\chi(h,h')=r_g\), and the corrected observable simplifies to

\begin{equation}
B_{\mathrm{corr}}=i\big(\chi(h)\chi(h')-c_\sigma(h,h')\mathbf1\big),
\qquad \varepsilon_\sigma(B_{\mathrm{corr}})=iss'.
\end{equation}

Although \(c_\sigma\) may be complex, the orthogonal-output formula is Hermitian under precisely this orthogonality condition, as follows already from the Hermitian commutator construction.

\subsection{Calibration within the attainable range}
\label{sec:calibration}

Let \(\mathcal H_{\mathrm{adm}}\) be the real linear space of allowed outgoing probe smearings for the chosen readout region, and let

\begin{equation}
\mathcal R_\lambda=F_\lambda(\mathcal H_{\mathrm{adm}})
\end{equation}

be its range in the system's physical CAR smearing space. The following construction requires two real-linearly independent physical elements of \(\mathcal R_\lambda\). This is a substantive range condition; compact support alone does not establish it.

Choose \(h_1,h_2\in\mathcal H_{\mathrm{adm}}\) with independent non-null images \(u=F_\lambda h_1\), \(v=F_\lambda h_2\). Define

\begin{equation}
\begin{gathered}
\alpha=\frac{r_\psi(u,v)}{\|u\|_\psi^2},\qquad
h=h_1,\qquad h'=h_2-\alpha h_1,\\
f_h=u,\qquad f_{h'}=v-\alpha u.
\end{gathered}
\end{equation}

Real linearity keeps the calibrated pair in the attainable range and, when both initial smearings lie in the same readout region, keeps their linear combination in that region. Directly,

\begin{equation}
r_\psi(f_h,f_{h'})=r_\psi(u,v)-\alpha\|u\|_\psi^2=0,
\end{equation}
\begin{equation}
\|f_{h'}\|_\psi^2
=\|v\|_\psi^2-\frac{r_\psi(u,v)^2}{\|u\|_\psi^2}>0.
\end{equation}

The strict inequality uses independence in the real physical CAR space. It can fail for dependent or null images. This argument establishes orthogonalization conditional on a two-dimensional attainable range, not access to arbitrary desired system modes. It also presupposes that real linear combinations are admissible experimental smearings; any further readout constraints must be checked separately.

\subsection{Normalization and the square of the induced observable}
\label{sec:normalization}

For the calibrated non-null pair, define

\begin{equation}
\begin{gathered}
n=\|f_h\|_\psi>0,\qquad n'=\|f_{h'}\|_\psi>0,\\
\Gamma=\frac{s}{n},\qquad\Gamma'=\frac{s'}{n'}.
\end{gathered}
\end{equation}

The CAR now imply

\begin{equation}
\Gamma^*=\Gamma,\quad \Gamma'^*=\Gamma',\quad
\Gamma^2=\Gamma'^2=\mathbf1,\quad
\Gamma\Gamma'=-\Gamma'\Gamma.
\end{equation}

Normalize the corrected Hermitian probe observable using the output norms:

\begin{equation}
\widehat B_\sigma(h,h')=\frac{B_{\mathrm{corr}}(h,h')}{nn'},
\qquad
A(h,h'):=\varepsilon_\sigma(\widehat B_\sigma)=i\Gamma\Gamma'.
\end{equation}

Hermiticity follows step by step:

\begin{equation}
A^*=-i(\Gamma\Gamma')^*
=-i\Gamma'\Gamma=i\Gamma\Gamma'=A.
\end{equation}

Its square is

\begin{equation}
\begin{aligned}
A^2&=i^2\Gamma\Gamma'\Gamma\Gamma'\\
&=-\Gamma(\Gamma'\Gamma)\Gamma'\\
&=\Gamma^2\Gamma'^2=\mathbf1.
\end{aligned}
\end{equation}

Consequently \(\operatorname{spec}(A)\subseteq\{-1,+1\}\), and \(P_\pm=(\mathbf1\pm A)/2\) are orthogonal projections. In a nonzero unital CAR representation containing both normalized modes, conjugation by \(\Gamma\) sends \(A\) to \(-A\), so both spectral values occur. The induced system bilinear is therefore dichotomic.

To distinguish this normalized dichotomy from the spectral properties of the probe bilinear, put \(a=\chi(h)\), \(b=\chi(h')\), \(u=\|h\|_\chi^2\), \(v=\|h'\|_\chi^2\), and \(q=r_\chi(h,h')\). Then \(a^2=u\mathbf1\), \(b^2=v\mathbf1\), and \(ba=-ab+2q\mathbf1\). Thus

\begin{equation}
(ab)^2=a(ba)b=-uv\mathbf1+2qab,
\end{equation}
\begin{equation}
\begin{aligned}
B_0^2&=-\big(ab-q\mathbf1\big)^2\\
&=-\big((ab)^2-2qab+q^2\mathbf1\big)\\
&=(uv-q^2)\mathbf1=\Delta^2\mathbf1,
\qquad \Delta=\sqrt{uv-q^2}.
\end{aligned}
\end{equation}

For independent physical probe modes, \(\Delta>0\), and the spectrum of \(B_0\) is \(\{-\Delta,+\Delta\}\). Indeed, \(a/\sqrt u\) is an involution whose conjugation sends \(B_0\) to \(-B_0\), so both values occur in a nonzero unital CAR representation. Since

\begin{equation}
\widehat B_\sigma=\frac{B_0-\beta_\sigma\mathbf1}{nn'},
\end{equation}

the corrected probe observable still has two spectral values,

\begin{equation}
\operatorname{spec}(\widehat B_\sigma)
=\left\{\frac{-\Delta-\beta_\sigma}{nn'},
\frac{\Delta-\beta_\sigma}{nn'}\right\},
\end{equation}

and its square is

\begin{equation}
\widehat B_\sigma^2
=\frac{(\Delta^2+\beta_\sigma^2)\mathbf1-2\beta_\sigma B_0}
{n^2n'^2}.
\end{equation}

For \(\Delta>0\), this square equals \(\mathbf1\) precisely when \(\beta_\sigma=0\) and \(\Delta=nn'\). These conditions are not imposed by output orthogonality and normalization alone. If the probe modes are dependent, \(\Delta=0\) and the bilinear is degenerate; no two-value claim is made in that case.

The normalized dichotomy proved for \(A\) therefore concerns the induced algebra element. The corrected probe observable is generally not an involution, although it has two spectral values for independent modes, and the induced map does not transfer its square multiplicatively. In particular,

\begin{equation}
\varepsilon_\sigma(\widehat B_\sigma^2)
\ne\varepsilon_\sigma(\widehat B_\sigma)^2
\end{equation}

in general. Equality of the induced first moment with \(A\) is not by itself an implementation of its sharp spectral measurement or of binary probe outcomes. An operational claim about a Bell experiment requires a specified readout POVM/instrument and a separate treatment of its statistics and probe noise \cite{FV2020}.

The conditions under which the corrected probe observable is itself an involution have a dynamical reading. By the scattering CAR identity \eqref{eq:carid}, $n^2=\|f_h\|_\psi^2=\|h\|_\chi^2-\|g_h\|_\chi^2$, so $\Delta=nn'$ holds precisely when $g_h=g_{h'}=0$, i.e.\ when both probe modes are transferred completely to the system; in that case also $c_\sigma=0$ and hence $\beta_\sigma=0$, so that $\widehat B_\sigma^{\,2}=\mathbf 1$ and the induced effects $(\mathbf 1\pm A)/2$ are obtained from the probe projections $(\mathbf 1\pm\widehat B_\sigma)/2$ without any first-moment restriction. Whether complete transfer is attainable depends on the form of the coupling. The mass-mixing density $\bar\psi\chi$ of Eq.~\eqref{eq:action} does not commute with the free dynamics, and the transfer it produces depends on the mode and on the duration of the coupling. For equal masses, the bilinear $i\psi^{T}\chi$, which is the charge density of the Dirac field $(\psi+i\chi)/\sqrt2$, generates an internal $SO(2)$ rotation of $(\psi,\chi)$ commuting with the free Hamiltonian; a coupling $\lambda k\,i\psi^{T}\chi$ with spatially constant profile and time-integrated strength $\pi/2$ then exchanges the two fields mode by mode, giving $g_h=0$ exactly. This is the fermionic counterpart of the rotation and swap protocols of Ref.~\cite{Fewster2026}. The construction of the present article does not depend on this choice; we note it because it identifies the coupling for which the first-moment identities of Sections~\ref{sec:construction}--\ref{sec:normalordering} become identities for the readout statistics. A detailed treatment is left to future work.

The construction can be summarized as follows:

\begin{equation*}
\boxed{\begin{gathered}
\text{free probe observable}\;\longrightarrow\;
\text{FV output map}\\[0.3em]
\longrightarrow\;\text{calibration}\;\longrightarrow\;
\text{normalization}\\[0.3em]
\longrightarrow\;\text{dichotomic induced observable}
\end{gathered}}
\end{equation*}

\section{Probe normal ordering and construction of quartic fermionic operators}
\label{sec:normalordering}
Let \(\tau^-\) be the identification of the incoming uncoupled algebra with the coupled algebra, and \(\tau^+\) the outgoing identification, with \(\Theta=(\tau^-)^{-1}\tau^+\).\footnote{The maps \(\tau^\pm\) are identification isomorphisms from the free, uncoupled system--probe algebra to the coupled algebra, determined by comparison in regions before and after the compact interaction and by the time-slice property. Thus \(\tau^-\) identifies the incoming description and \(\tau^+\) the outgoing description; \(\Theta=(\tau^-)^{-1}\tau^+\) reexpresses outgoing observables in the incoming algebra.} Denote the outgoing coupled field by

\begin{equation}
Y_h:=\tau^+(\mathbf1\widehat\otimes\chi(h)),
\qquad (\tau^-)^{-1}(Y_h)=X_h=p+s.
\end{equation}

Thus \(Y_h\), \(X_h\), and \(\chi(h)\) are related but live in different algebraic descriptions. In the outgoing notation used in the manuscript, \(\chi_{\mathrm{out}}(h)\) denotes \(Y_h\); equations using its incoming decomposition denote its pullback \(X_h\).

Define

\begin{equation}
\begin{gathered}
:\!Y_hY_{h'}\!:\!_{\mathrm{probe},\sigma}
:=Y_hY_{h'}-c_\sigma(h,h')\mathbf1,\\
c_\sigma(h,h')=\sigma\big(\chi_{\mathrm{in}}(g_h)
\chi_{\mathrm{in}}(g_{h'})\big).
\end{gathered}
\end{equation}

For \(r_\psi(f_h,f_{h'})=0\), \(n=\|f_h\|_\psi>0\), and \(n'=\|f_{h'}\|_\psi>0\), the normalized two-field prescription is

\begin{equation}
\widetilde B_\sigma(h,h')
:=\frac{i}{nn'}:\!Y_hY_{h'}\!:\!_{\mathrm{probe},\sigma}.
\end{equation}

The corresponding abstract probe observable, to which the FV map is applied, is

\begin{equation}
\boxed{\begin{gathered}
\widehat B_\sigma(h,h')
=\frac{i}{nn'}\big(\chi(h)\chi(h')-c_\sigma(h,h')\mathbf1\big),\\
\widetilde B_\sigma=\tau^+(\mathbf1\widehat\otimes\widehat B_\sigma).
\end{gathered}}
\end{equation}

Its Hermiticity can also be checked without suppressing the subtraction. Put \(q=r_\chi(h,h')\). By the scattering CAR identity and output orthogonality, \(q=r_g\); by the contraction identity, \(c_\sigma+\overline c_\sigma=2q\). Therefore

\begin{equation}
\begin{aligned}
\widehat B_\sigma^*
&=\frac{-i}{nn'}\big(\chi(h')\chi(h)-\overline c_\sigma\mathbf1\big)\\
&=\frac{-i}{nn'}\big(-\chi(h)\chi(h')+(2q-\overline c_\sigma)\mathbf1\big)\\
&=\frac{i}{nn'}\big(\chi(h)\chi(h')-c_\sigma\mathbf1\big)
=\widehat B_\sigma.
\end{aligned}
\end{equation}

The normalized induced two-field identity is

\begin{equation}
\begin{aligned}
\eta_\sigma\big((\tau^-)^{-1}(\widetilde B_\sigma)\big)
&=\varepsilon_\sigma(\widehat B_\sigma)\\
&=\frac{i\psi(f_h)\psi(f_{h'})}{nn'}=A(h,h').
\end{aligned}
\end{equation}

If \(\langle\cdot\rangle_\sigma\) is retained in place of \(\eta_\sigma\), it must be defined as a probe-only expectation after incoming pullback, leaving a system algebra element. It is not the full scalar expectation in the coupled state. The full expectation is \((\omega\widehat\otimes\sigma)((\tau^-)^{-1}(\widetilde B_\sigma))=\omega(A)\). Hermiticity and the identity \(A^2=\mathbf1\) then follow from the explicit steps of the normalized CAR and square calculation.

If output orthogonality is not satisfied, simply multiplying the subtracted product by \(i\) is insufficient. The Hermitian commutator construction, followed by the required calibration, supplies the correct formulation. Neither version assumes multiplicativity of \(\varepsilon_\sigma\) or arbitrary access to system smearings.

For the Alice--Bob four-field expression the same prescription is crucial. Schematically, with independent local probes,
\begin{equation}
\boxed{\begin{aligned}
&\left\langle:\!\chi_{B,\mathrm{out}}(b)\chi_{B,\mathrm{out}}(b')
\chi_{A,\mathrm{out}}(a)\chi_{A,\mathrm{out}}(a')\!:\!_\chi\right\rangle_\sigma\\
&\qquad=\psi(f_b)\psi(f_{b'})\psi(f_a)\psi(f_{a'}),
\end{aligned}}
\end{equation}
with the ordering fixed consistently with the fermionic convention. Probe-only contractions, including the unwanted product of probe two-point functions, are removed by the probe normal ordering. The system factors are not normal ordered away.

To connect this quartic expression with Sec.~\ref{sec:bell}, let \(N_A=\|f_a\|_\psi\|f_{a'}\|_\psi\) and \(N_B=\|f_b\|_\psi\|f_{b'}\|_\psi\), with non-null, real-orthogonal induced smearings within each local pair, and define \(A_a=i\psi(f_a)\psi(f_{a'})/N_A\) and \(B_b=i\psi(f_b)\psi(f_{b'})/N_B\). Incorporating the factor \(i\) for each bilinear and dividing the four-field expression by \(N_BN_A\) gives \(-\psi(f_b)\psi(f_{b'})\psi(f_a)\psi(f_{a'})/(N_BN_A)=B_bA_a\), since \(i^2=-1\). If the induced supports lie in spacelike separated regions, graded locality makes these even bilinears commute, so this is also \(A_aB_b\). Here \(a'\) and \(b'\) label the second smearing within each bilinear, not the alternative measurement settings denoted by \(A'\) and \(B'\) in Sec.~\ref{sec:bell}. Under the independent even probe preparation and the probe-only contraction-subtraction prescription underlying the displayed quartic identity, its normalized system expectation is therefore \(\omega(A_aB_b)\), one of the correlations entering the Bell--CHSH combination. This relation is obtained from the expanded joint two-probe expression, not from multiplicativity of \(\varepsilon_\sigma\); it neither establishes a sharp binary readout nor demonstrates a Bell violation.

\section{Induced Bell--CHSH structure}
\label{sec:bell}
Choose two calibrated pairs for Alice and two for Bob, producing dichotomic observables
\begin{equation}
A,A',B,B',\qquad A^2=A'^2=B^2=B'^2=\mathbf1.
\end{equation}
Their local interaction profiles determine the localization of the induced smearings through the support inclusion stated in Sec.~\ref{sec:intro}. The CHSH operator is
\begin{equation}
\mathcal B=A(B+B')+A'(B-B').
\end{equation}
Thus the FV construction reduces the operational problem to engineering, through the probe smearings, the four induced Majorana bilinears required by the chosen Bell configuration. The complicated spacetime dependence carried by the exact M\o ller solution is contained in the functions $b_{\lambda,h}$ and need not be explicitly evaluated for the structural construction.

\subsection{Conditional maximal Bell--CHSH expectation}
\label{sec:maximal}
Individual induced involutions do not by themselves supply the incompatible local settings needed for maximal CHSH correlations. Assume, in addition, that each local output map has three independent non-null real modes with spacelike separated induced representatives, and that real linear calibration is admissible. Appendix~\ref{app:pauli} constructs from their normalized Majorana operators the even Pauli triples $(X_A,Y_A,Z_A)$ and $(X_B,Y_B,Z_B)$, with anticommutation within each triple and commutation between the two sides. The settings $A=Z_A$, $A'=X_A$, $B=-(Z_B+X_B)/\sqrt2$ and $B'=-(Z_B-X_B)/\sqrt2$ are attainable bilinears within these assumed ranges.

The singlet projection $P_s=(\mathbf1-X_AX_B-Y_AY_B-Z_AZ_B)/4$ gives a normalized singlet vector from the vacuum whenever its projection is nonzero. If that projection vanishes, Appendix~\ref{app:pauli} supplies an even Pauli rotation of the vacuum with a nonzero singlet projection, without assuming vacuum cyclicity. In this conditionally prepared system state $\omega_s$, the correlations are $\omega_s(A(\mathbf a)B(\mathbf b))=-\mathbf a\cdot\mathbf b$, and the above settings give
\begin{equation}
\boxed{\omega_s(\mathcal B)=2\sqrt2,\qquad
|\omega(\mathcal B)|\leq2\sqrt2.}
\end{equation}
Thus the induced operator construction can saturate the Tsirelson bound under the stated range, localization and state-preparation hypotheses \cite{Tsirelson1980}. This is not a claim of maximal violation in the original vacuum, nor a proof of an operational binary probe measurement: those require the additional readout and preparation conditions discussed in Appendix~\ref{app:pauli}.

\section{Discussion and Conclusion}
\label{sec:discussion}
Three points are central. First, localization of the induced system observable follows from the exact factorization $f_h=kb_{\lambda,h}$. Second, dichotomy is obtained at the level of the induced output observable by imposing real orthogonality and then normalizing the Majorana bilinear. Third, probe normal ordering removes the background probe contractions that otherwise contaminate the four-field Bell expression. Together these observations provide a direct route from local FV probe couplings to induced Bell--CHSH observables.

Relative to the Summers--Werner theorem \cite{SummersWerner1985,SummersWerner1987}, the construction replaces an existence statement by explicit objects: the local observables are induced bilinears $A(h,h')=\varepsilon_\sigma(\widehat B_\sigma)$ whose smearings are fixed by the coupling profile and the probe readout, their localization follows from the factorization rather than from the modular structure of the wedge algebras, and the maximal correlation is attained in a state prepared from the vacuum by an explicit even projection. The relation between FV-realizable operations and Bell correlations has recently been examined from a complementary angle by Mandrysch, Simmons and Navascu\'es~\cite{Mandrysch2026}. They consider a Bell configuration in which a channel with compact interaction zone is placed between setting-dependent local operations in the causal past and fixed local measurements in the causal future, and show that whenever this intermediate channel admits an FV realization, the resulting correlations lie in the commuting-operator quantum set. Their argument relies only on the scattering morphism being a causal automorphism together with microcausality, and so carries over to the graded setting considered here once attention is restricted to even observables. Our construction occupies a different slot of that scheme: the FV couplings implement the local observables themselves, the coupling regions $K_A\subset W_R$ and $K_B\subset W_L$ are spacelike separated, and no non-trivial inter-party channel is present. The causal factorization $\Theta=\Theta_A\circ\Theta_B$ of Eq.~\eqref{eq:alice-bob-factorization} is precisely the mechanism that, in~\cite{Mandrysch2026}, yields the cross-side commutation underlying the quantum bound. The two results are thus consistent and complementary: Ref.~\cite{Mandrysch2026} bounds what an FV channel can mediate between two parties, while Appendix~\ref{app:pauli} exhibits FV-induced observables and a conditionally prepared state that saturate that bound. Ref.~\cite{Mandrysch2026} also constructs causal channels for free scalar fields whose correlations fall outside the quantum set and which therefore admit no FV realization; the present construction does not address this question.

The factorization separates the support of an induced smearing representative from the dynamical information contained in $b_{\lambda,h}$. Appendix~\ref{app:moller} explains why the final interaction vertex in the dual M\o ller expansion supplies a factor of $k$ at every order. Under the assumed well-defined M\o ller map and the conditions needed for exact resummation, this yields the stated support inclusion without requiring a closed expression for the scattering transformation. A merely formal series gives the corresponding order-by-order statement. The support control is therefore a structural localization result, not a characterization of the range of $F_\lambda$: it does not imply that arbitrary system modes supported in the interaction region can be reached by allowed probe smearings.

Calibration addresses a separate issue within that attainable range. Given two independent non-null real physical smearing classes and admissible real linear combinations of their probe preimages, Gram--Schmidt orthogonalization produces a real-orthogonal output pair without leaving the range. Normalization in the physical CAR norm then turns its Hermitian Majorana bilinear into an involution. The required independence and non-nullness are substantive hypotheses, and further restrictions on experimental smearings must be checked separately. This establishes dichotomy for the induced system algebra element; it does not establish that the corrected probe observable is a normalized involution or that its readout implements a sharp binary measurement of that element. As noted at the end of Sec.~\ref{sec:normalization}, the latter is obtained exactly for couplings that transfer the probe modes completely, which for equal masses is achieved by a coupling to the inter-species charge density $i\psi^{T}\chi$ rather than to the mass-mixing density $\bar\psi\chi$.

Probe normal ordering supplies the corresponding control of correlations. The subtraction acts exclusively on contractions in the prepared probe state, leaving the system factors intact. For independent even local probes, the expanded joint four-field expression, with the stated ordering and normalization, yields the product of the induced bilinears whose system expectation enters CHSH. Spacelike separation makes these even Alice--Bob observables commute. This argument uses the joint expansion before taking the probe expectation: although the scattering isomorphism is multiplicative, the induced-observable map generally is not. Consequently, the two-field first-moment identities alone cannot be multiplied to infer joint readout statistics.

Appendix~\ref{app:pauli} adds the hypotheses needed to pass from individual induced involutions to a maximal CHSH expectation. Three independent non-null attainable modes on each side, with admissible real calibration and spacelike separated representatives, generate two commuting even Pauli algebras and supply incompatible local settings within those ranges. The singlet construction uses a nonzero joint projection of the vacuum, preceded when necessary by an even Pauli rotation. It is conditional state preparation, and the article does not show that the FV readout implements this projection. In the resulting prepared system state the selected attainable settings saturate the Tsirelson bound $2\sqrt2$; this is neither a claim of violation in the original vacuum nor, by itself, a realization of a binary-outcome probe experiment.

In summary, the construction identifies the algebraic ingredients for Bell--CHSH observables while keeping localization, attainability, normalization and state preparation explicit, and it places them where the Summers--Werner theorem leaves them implicit: in the smearings, the couplings and the probe readouts. Two steps separate these induced correlations from an implemented protocol. The first is to characterize the ranges of the local output maps for specified interaction profiles and admissible readout regions, including whether they contain the independent modes required by (A3). The second is the readout itself: the probe POVMs or instruments, their joint statistics and outcome normalization, probe noise and the conditional preparation. For the second step the analysis of Sec.~\ref{sec:normalization} singles out a definite direction. For equal masses the inter-species charge density $i\psi^{T}\chi$ generates an internal rotation commuting with the free dynamics, and a coupling of time-integrated strength $\pi/2$ transfers each probe mode completely, so that $g_h=0$, $c_\sigma=0$ and the corrected probe observable is itself an involution. The first-moment identities of Secs.~\ref{sec:construction}--\ref{sec:normalordering} then become identities for the readout statistics, and the Bell test is performed by sharp binary probe readouts. Working this out for compactly supported profiles, with the attainable ranges determined for $K_A$ and $K_B$, is the natural continuation of the present work.

\begin{acknowledgments}
It is a pleasure to thank Christopher J. Fewster for a lucid presentation of the measurement framework to our group in June 2026. The authors thank the Brazilian agencies CNPq, CAPES and FAPERJ for financial support. S.~P.~Sorella, I.~Roditi and M.~S.~Guimaraes are CNPq researchers under contracts 302991/2024-7, 319060/2025-0 and 309793/2023-8, respectively.

The authors acknowledge the use of AI assistants (ChatGPT, OpenAI, and Claude, Anthropic) for exploratory calculations, numerical cross-checks and language editing; all scientific content, calculations, interpretations and conclusions were checked by the authors, who remain solely responsible for them.
\end{acknowledgments}

\appendix
\section{Exact factorization from the iterative M\o ller solution}
\label{app:moller}
Let
\begin{equation}
D_\psi\psi+\lambda k\chi=0,\qquad D_\chi\chi+\lambda k\psi=0,
\end{equation}
where $D_j=i\gamma^\mu\partial_\mu-m_j$ are the real operators fixed in Sec.~\ref{sec:model}. For the incoming solution choose the retarded Green operators $S_\psi^{\mathrm{ret}},S_\chi^{\mathrm{ret}}$ and free incoming fields $\psi_0,\chi_0$. The equations derived from the complete action give
\begin{equation}
\psi=\psi_0-\lambda S_\psi^{\mathrm{ret}}(k\chi),\qquad
\chi=\chi_0-\lambda S_\chi^{\mathrm{ret}}(k\psi).
\end{equation}
Applying $D_jS_j^{\mathrm{ret}}=\mathbf1$ verifies the minus signs in these integral equations. The outgoing construction uses advanced Green operators with the same source signs. No additional $i$, $C$ or $M$ is inserted at a mixing vertex: after multiplication by $(\gamma^0)^{-1}$ in the variation, that vertex is the real scalar operator $\lambda k\mathbf1_2$.

Iterating gives the Neumann/M\o ller series, with these signs and the distinct masses contained in its coefficients. The series interpretation assumes convergence on the relevant test-function space; if used only formally, it is an order-by-order statement, and an exact resummation requires the well-defined M\o ller map assumed in Sec.~\ref{sec:model}. When the outgoing probe field is smeared with $h$ and the propagators are transferred to the test-function side, the system component contains only odd powers of the coupling. The first terms have the structure
\begin{align}
f_h^{(1)}&=k\Lambda_h^{(1)},\\
f_h^{(3)}&=k\Lambda_h^{(3)},\\
f_h^{(5)}&=k\Lambda_h^{(5)},
\end{align}
and inductively
\begin{equation}
f_h^{(2n+1)}(x)=k(x)\Lambda_h^{(2n+1)}(x).
\end{equation}
The reason is elementary: on the dual test-function map, the final interaction vertex connecting the probe chain to the system is multiplication by $k$. All propagators and all earlier interaction insertions are contained in $\Lambda_h^{(2n+1)}$.

Resumming the exact linear solution,
\begin{equation}
f_h(x)=\sum_{n\geq0}f_h^{(2n+1)}(x)=k(x)\sum_{n\geq0}\Lambda_h^{(2n+1)}(x).
\end{equation}
Define
\begin{equation}
b_{\lambda,h}(x):=\sum_{n\geq0}\Lambda_h^{(2n+1)}(x).
\end{equation}
Then
\begin{equation}
f_h(x)=k(x)b_{\lambda,h}(x),
\end{equation}
which is the exact structural form used in the main text. In particular,
\begin{equation}
\supp f_h\subseteq\supp k.
\end{equation}
No explicit closed expression for the M\o ller operator is required for this conclusion.

\section{Majorana Pauli algebras, singlet preparation and maximal CHSH expectation}
\label{app:pauli}
\subsection{Attainable local modes and their even Pauli algebras}
Assume that the real ranges of the local output maps $F_A,F_B$ each contain at least three independent non-null physical smearing classes, represented in spacelike separated interaction regions. Real Gram--Schmidt within each admissible readout space then gives $e_{L1},e_{L2},e_{L3}$, $L=A,B$, with $r_\psi(e_{Li},e_{Lj})=\delta_{ij}$. Graded locality gives $r_\psi(e_{Ai},e_{Bj})=0$. These statements require attainable representatives with the stated localization; they do not follow merely from the support inclusion or from individual dichotomy. Define
\begin{equation}
\begin{gathered}
\gamma_{Li}=\psi(e_{Li}),\qquad
\gamma_{Li}^*=\gamma_{Li},\\
\{\gamma_{Li},\gamma_{Mj}\}=2\delta_{LM}\delta_{ij}\mathbf1.
\end{gathered}
\end{equation}
The symbol $\gamma_{Li}$ here denotes a smeared Majorana operator, not a Dirac matrix $\gamma^\mu$ of Sec.~\ref{sec:model}. Put
\begin{equation}
X_L=-i\gamma_{L2}\gamma_{L3},\quad
Y_L=-i\gamma_{L3}\gamma_{L1},\quad
Z_L=-i\gamma_{L1}\gamma_{L2}.
\label{eq:majorana-pauli}
\end{equation}
The minus signs are orientation choices: reversing a smearing pair writes each generator in the $+i$ convention of Sec.~\ref{sec:construction}. Each is Hermitian and squares to $\mathbf1$. For example,
\begin{align}
X_L^*&=i\gamma_{L3}\gamma_{L2}=-i\gamma_{L2}\gamma_{L3}=X_L,\\
X_L^2&=-\gamma_{L2}\gamma_{L3}\gamma_{L2}\gamma_{L3}=\mathbf1,\\
X_LY_L&=-\gamma_{L2}\gamma_{L3}\gamma_{L3}\gamma_{L1}
=\gamma_{L1}\gamma_{L2}=iZ_L,\\
Y_LX_L&=-\gamma_{L3}\gamma_{L1}\gamma_{L2}\gamma_{L3}
=-\gamma_{L1}\gamma_{L2}=-iZ_L.
\end{align}
Cyclic permutations similarly give $Y_LZ_L=iX_L$ and $Z_LX_L=iY_L$, with opposite signs in the reversed products. Thus, writing $\boldsymbol\sigma_L=(X_L,Y_L,Z_L)$,
\begin{equation}
\sigma_{Li}\sigma_{Lj}=\delta_{ij}\mathbf1
+i\sum_k\epsilon_{ijk}\sigma_{Lk},\qquad
\{\sigma_{Li},\sigma_{Lj}\}=2\delta_{ij}\mathbf1.
\end{equation}
A bilinear on Alice's side commutes with one on Bob's side: moving two odd factors past two others contributes $(-1)^4=+1$. Hence $[\sigma_{Ai},\sigma_{Bj}]=0$. This constructs incompatibility within each side as well as compatibility between sides, rather than inferring them from the square of an individual observable.

The unital even algebra generated by each triple is $M_2(\mathbb C)$, and the two commuting copies generate $M_2\otimes M_2\simeq M_4(\mathbb C)$. Their representation is faithful because a nonzero unital representation of this simple matrix algebra has zero kernel. On any nonzero parity sector its Hilbert-space representation therefore has the form $(\mathbb C^2\otimes\mathbb C^2)\otimes\mathcal K$, with the Pauli generators acting on the first two factors. All generators are even and preserve total fermion parity. This is a factorization of the finite even observable algebra; it does not assign a separate Fock space to three odd modes on each side. Six total Majorana modes give a consistent CAR representation, and no fourth local mode is required for these even Pauli algebras.

For unit vectors $\mathbf a,\mathbf b\in\mathbb R^3$, set $A(\mathbf a)=\mathbf a\cdot\boldsymbol\sigma_A$ and $B(\mathbf b)=\mathbf b\cdot\boldsymbol\sigma_B$. The Pauli identity implies $A(\mathbf a)^2=B(\mathbf b)^2=\mathbf1$. The concrete settings used below remain single Majorana bilinears, with rotated real smearings inside the same attainable three-dimensional range; no assumption of surjectivity onto arbitrary field modes is needed.

\subsection{A nonzero singlet vector obtained from the vacuum}
Let $\Omega$ be a normalized even vacuum vector in a nonzero Hilbert-space representation of the above algebra. No cyclicity or separating property is needed for the finite-algebra argument below. Define
\begin{equation}
K=X_AX_B+Y_AY_B+Z_AZ_B,\qquad
P_s=\frac14(\mathbf1-K).
\label{eq:singlet-projector}
\end{equation}
Products on different sides commute. The diagonal terms of $K^2$ give $3\mathbf1$; each ordered off-diagonal product is
$(\sigma_{Ai}\sigma_{Aj})(\sigma_{Bi}\sigma_{Bj})=-\sigma_{Ak}\sigma_{Bk}$ for the remaining index $k$. Each $k$ occurs twice, so $K^2=3\mathbf1-2K$. Consequently
\begin{equation}
\begin{gathered}
P_s^*=P_s,\\
P_s^2=\frac1{16}(\mathbf1-2K+K^2)=\frac14(\mathbf1-K)=P_s.
\end{gathered}
\end{equation}
In the faithful Pauli representation choose $Z|0\rangle=|0\rangle$, $Z|1\rangle=-|1\rangle$ and $X|0\rangle=|1\rangle$. Directly,
\begin{equation}
\begin{gathered}
|s\rangle=\frac{|01\rangle-|10\rangle}{\sqrt2},\qquad
P_s=|s\rangle\langle s|\otimes\mathbf1_{\mathcal K},\\
(\sigma_{Ai}+\sigma_{Bi})|s\rangle=0.
\end{gathered}
\end{equation}
If $q_0=\langle\Omega,P_s\Omega\rangle>0$, the desired vector is $\Omega_s=P_s\Omega/\sqrt{q_0}$. A direct projection of the vacuum can, however, be zero. To cover that case explicitly, set
\begin{equation}
\begin{gathered}
U_0=\mathbf1,\qquad U_1=X_B,\qquad U_2=Y_B,\qquad U_3=Z_B,\\
P_\mu=U_\mu P_sU_\mu^*,\qquad
q_\mu=\langle\Omega,P_\mu\Omega\rangle.
\end{gathered}
\end{equation}
The vectors $(\mathbf1\otimes\sigma_\mu)|s\rangle$, with $\sigma_0=\mathbf1$, are the four orthonormal Bell vectors: their inner products are $\frac12\operatorname{tr}(\sigma_\mu^*\sigma_\nu)=\delta_{\mu\nu}$. Therefore $\sum_{\mu=0}^3P_\mu=\mathbf1$ and $\sum_\mu q_\mu=1$. At least one $q_\mu$ is positive, in fact at least $1/4$. For any such index the concrete construction is
\begin{equation}
\begin{gathered}
\Omega_s=\frac{P_sU_\mu^*\Omega}{\sqrt{q_\mu}},\\
\|P_sU_\mu^*\Omega\|^2=q_\mu>0,\qquad P_s\Omega_s=\Omega_s.
\end{gathered}
\label{eq:vacuum-singlet}
\end{equation}
Both $U_\mu$ and $P_s$ are even, so $\Omega_s$ has the vacuum's total parity. This constructs a singlet restriction to the two Pauli algebras from a nonzero vector in the vacuum's even algebraic orbit. It does not assert that the vacuum itself is a singlet. Although $U_\mu$ is a local even unitary, $P_s$ is a joint projection: the construction is conditional state preparation, not a deterministic preparation by local unitaries, and the FV readout scheme has not been shown to implement this projection.

\subsection{Singlet correlations and an attainable choice of settings}
Define $\omega_s(T)=\langle\Omega_s,T\Omega_s\rangle$. Since the first two representation factors are in $|s\rangle$, their single-side Pauli expectations vanish. The relation $(\sigma_{Ai}+\sigma_{Bi})|s\rangle=0$ and the Pauli product identity give
\begin{equation}
\begin{gathered}
\omega_s(\sigma_{Ai}\sigma_{Bj})
=-\langle s|\sigma_{Ai}\sigma_{Aj}|s\rangle
=-\delta_{ij},\\
\omega_s(A(\mathbf a)B(\mathbf b))=-\mathbf a\cdot\mathbf b.
\end{gathered}
\end{equation}
Choose
\begin{equation}
\begin{gathered}
A=Z_A,\qquad A'=X_A,\\
B=-\frac{Z_B+X_B}{\sqrt2},\qquad
B'=-\frac{Z_B-X_B}{\sqrt2}.
\end{gathered}
\label{eq:optimal-settings}
\end{equation}
Anticommutation of $X_L,Z_L$ gives $A^2=A'^2=B^2=B'^2=\mathbf1$. In particular the Bob settings are not merely formal sums of unattainable observables:
\begin{equation}
B=i\frac{\gamma_{B1}-\gamma_{B3}}{\sqrt2}\gamma_{B2},\qquad
B'=i\frac{\gamma_{B1}+\gamma_{B3}}{\sqrt2}\gamma_{B2}.
\end{equation}
The rotated mode pairs are real orthonormal; their probe preimages are the corresponding real linear combinations of the three assumed admissible preimages. Alice's settings are likewise the oriented pairs $i\gamma_{A2}\gamma_{A1}$ and $i\gamma_{A3}\gamma_{A2}$. Each setting is thus an induced bilinear of Sec.~\ref{sec:construction} under the stated range hypothesis.

For the CHSH sign convention in Sec.~\ref{sec:bell}, the four correlations are
\begin{equation}
\begin{gathered}
\omega_s(AB)=\omega_s(AB')=\omega_s(A'B)=\frac1{\sqrt2},\\
\omega_s(A'B')=-\frac1{\sqrt2}.
\end{gathered}
\end{equation}
Equivalently,
\begin{equation}
\begin{gathered}
\mathcal B=-\sqrt2(Z_AZ_B+X_AX_B),\qquad
\mathcal B\Omega_s=2\sqrt2\,\Omega_s,\\
\omega_s(\mathcal B)=2\sqrt2.
\end{gathered}
\end{equation}
Reversing both Bob settings gives $-2\sqrt2$. These expectations exceed the local hidden-variable bound $2$ for the normalized CHSH expression \cite{CHSH1969}.

\subsection{Maximality and the scope of the induced result}
For arbitrary Hermitian involutions $A,A',B,B'$ with all Alice--Bob commutators zero, the two squared summands satisfy
\begin{align}
[A(B+B')]^2&=2\mathbf1+\{B,B'\},\\
[A'(B-B')]^2&=2\mathbf1-\{B,B'\}.
\end{align}
The cross terms are $-AA'[B,B']+A'A[B,B']=-[A,A'][B,B']$, hence
\begin{equation}
\begin{gathered}
\mathcal B^2=4\mathbf1-[A,A'][B,B'],\\
\|[A,A']\|\leq2,\qquad\|[B,B']\|\leq2.
\end{gathered}
\end{equation}
Since $\mathcal B$ is Hermitian, $\|\mathcal B\|^2=\|\mathcal B^2\|\leq8$, and every normalized state obeys $|\omega(\mathcal B)|\leq2\sqrt2$. The singlet settings above saturate this Tsirelson bound \cite{Tsirelson1980}. The elementary bound requires cross-side commutation and normalization; individual dichotomy alone does not imply saturation.

The result is an exact expectation in a conditionally prepared system state for attainable induced algebra elements. It does not use multiplicativity of the FV induced map. Identifying these correlations with an actual probe experiment additionally requires suitable joint readout statistics, normalized binary outcomes or an equivalent specified POVM/instrument, and control of probe noise and preparation. The normal-ordered first-moment identities of this article alone establish none of these operational requirements.

\section{Generalization to bosonic systems}
\label{app:bosonic}
The sign--cosine construction for scalar fields provides a possible bosonic counterpart to the Majorana bilinears \cite{FiniteWeyl2026}. The prescription considered here expands the sign of the cosine in odd Fourier harmonics and normally orders only the incoming probe contribution in each harmonic. In the probe vacuum, its expectation is then unity, leaving the corresponding system Weyl operator. This cancellation is exact for every harmonic and every finite Fourier polynomial. We formulate it below and show that the induced finite approximants converge to a bounded system sign--cosine observable. We distinguish this statement from convergence of the normally ordered probe series itself, which is a separate requirement for interpreting the infinite expression as an FV readout observable.

\subsection{Coupled Klein--Gordon fields and the induced Weyl map}
In this appendix only, $\phi$ denotes the massive real scalar system field and $\psi$ a massive real scalar probe, replacing the fermionic fields of the main text. Adopt the action
\begin{equation}
\begin{aligned}
S_{\mathrm b}={}&\frac12\int d^2x\,\big[
\partial_\mu\phi\partial^\mu\phi-m_\phi^2\phi^2\\
&\qquad+\partial_\mu\psi\partial^\mu\psi-m_\psi^2\psi^2\big]\\
&-\lambda\int d^2x\,k\phi\psi.
\end{aligned}
\end{equation}
This sign convention gives $P\phi+\lambda k\psi=0$ and $Q\psi+\lambda k\phi=0$, where $P=\Box+m_\phi^2$, $Q=\Box+m_\psi^2$, and $k$ is real and compactly supported. Assume well-defined advanced and retarded Green operators for
\begin{equation}
T=\begin{pmatrix}P&\lambda k\\\lambda k&Q\end{pmatrix},
\qquad V=T-(P\oplus Q).
\end{equation}
For a real outgoing probe test function $h$ supported in the region after the coupling, the FV incoming pullback is \cite{FV2020}
\begin{align}
\binom{f_h}{g_h}&=\binom{0}{h}-V E_T^{\mathrm{adv}}\binom{0}{h},\\
\Theta_{\mathrm b}(\mathbf1\otimes\psi(h))
&=\phi(f_h)\otimes\mathbf1+\mathbf1\otimes\psi(g_h).
\label{eq:bosonic-pullback}
\end{align}
In particular $f_h=-\lambda k(E_T^{\mathrm{adv}}(0,h)^T)_\psi$ has a compactly supported representative in the coupling region. These are identities of smeared fields and physical smearing classes, not pointwise decompositions with an arbitrary prescribed system function. The bosonic tensor product is ungraded. For independent local probes, each copy has its own such maps; the compact right- and left-wedge couplings of Section~\ref{sec:alice-bob} retain causal factorization.

Write $W_\phi(f)=e^{i\phi(f)}$ and adopt
\begin{equation}
\begin{gathered}
[\phi(f),\phi(f')]=iE_\phi(f,f')\mathbf1,\\
W_\phi(f)W_\phi(f')=e^{-iE_\phi(f,f')/2}W_\phi(f+f'),
\end{gathered}
\label{eq:bosonic-weyl}
\end{equation}
with analogous probe conventions. Scattering preserves the CCR, hence
\begin{equation}
E_\psi(h,h')=E_\phi(f_h,f_{h'})+E_\psi(g_h,g_{h'}).
\end{equation}
Prepare the incoming state as $\omega\otimes\sigma$, with a centered regular quasifree probe state satisfying
\begin{equation}
\sigma(W_\psi(g))=e^{-\mu_\sigma(g,g)/2},\qquad
v_h:=\mu_\sigma(g_h,g_h)\geq0.
\end{equation}
The exact induced map on bounded Weyl observables is
\begin{equation}
\varepsilon_\sigma(W_\psi(th))
=e^{-t^2v_h/2}W_\phi(tf_h),\qquad t\in\mathbb R.
\label{eq:bosonic-noise}
\end{equation}
The damping factor is the characteristic function of the residual incoming probe quadrature. It is not removed by linearity of scattering.

\subsection{The sign--cosine readout and the meaning of probe normal ordering}
Set $q(x)=\operatorname{sgn}(\cos x)$, taking $\operatorname{sgn}(0)=1$. In the regular scalar Fock representation nonzero quadratures have no spectral atoms at the zeros of cosine. Thus $q(\psi(h))$ is a bounded Hermitian involution. Let $q_N$ be its ordinary Fej\'er approximation,
\begin{equation}
\begin{gathered}
q_N(x)=\sum_{\substack{1\leq n\leq N\\n\ \mathrm{odd}}}
a_n\Big(1-\frac{n}{N+1}\Big)\cos(nx),\\
a_n=\frac4\pi\frac{(-1)^{(n-1)/2}}n.
\end{gathered}
\end{equation}
These polynomials obey $\|q_N\|_\infty\leq1$ and converge strongly under spectral functional calculus to $q$, subject to the preceding condition at the jump points.

For the probe-vacuum prescription, specialize $\sigma$ to the probe vacuum and let $v_h$ be its variance. Vacuum normal ordering of the incoming probe Weyl factor means
\begin{equation}
\begin{gathered}
:\!W_\psi(ng_h)\!:\!_{\sigma}
=e^{n^2v_h/2}W_\psi(ng_h),\\
\sigma(:\!W_\psi(ng_h)\!:\!_{\sigma})=1.
\end{gathered}
\end{equation}
Therefore the probe-only normally ordered outgoing harmonic, expressed after incoming pullback, satisfies the exact identity
\begin{equation}
\eta_\sigma\big(W_\phi(nf_h)\otimes:\!W_\psi(ng_h)\!:\!_{\sigma}\big)
=W_\phi(nf_h).
\label{eq:bosonic-ordered-weyl}
\end{equation}
To display the complete harmonic step, put $P_h=\psi(g_h)$ and $S_h=\phi(f_h)$ in their commuting incoming tensor factors. For $n=2r+1$,
\begin{align}
&:\!\cos(n(P_h+S_h))\!:\!_{\mathrm{probe},\sigma}\nonumber\\
&\quad=\frac12\left[e^{inS_h}:\!e^{inP_h}\!:\!_\sigma
+e^{-inS_h}:\!e^{-inP_h}\!:\!_\sigma\right],\\
&\eta_\sigma\big(:\!\cos(n(P_h+S_h))\!:\!_{\mathrm{probe},\sigma}\big)\nonumber\\
&\quad=\frac12(e^{inS_h}+e^{-inS_h})=\cos(nS_h).
\end{align}
Multiplication by $4(-1)^r/[\pi(2r+1)]$ and summation over the odd harmonics reproduces the Fourier recipe for the sign of the cosine, with only the system factor remaining. Probe-only Wick subtraction can thus be defined unambiguously for each finite harmonic by multiplying it by $e^{n^2v_h/2}$. Accordingly the bounded, but generally noncontractive, finite probe observable
\begin{equation}
B_{N,h}=\sum_{\substack{1\leq n\leq N\\n\ \mathrm{odd}}}
a_n\Big(1-\frac{n}{N+1}\Big)e^{n^2v_h/2}\cos(n\psi(h))
\end{equation}
satisfies the exact first-moment identity
\begin{equation}
\varepsilon_\sigma(B_{N,h})=q_N(\phi(f_h)).
\label{eq:bosonic-finite-subtraction}
\end{equation}
In particular, with the spectral assumptions stated above,
\begin{equation}
\operatorname*{s-lim}_{N\to\infty}
\varepsilon_\sigma(B_{N,h})=q(\phi(f_h)).
\label{eq:bosonic-induced-limit}
\end{equation}
Equivalently, writing $P_h=\psi(g_h)$ and $S_h=\phi(f_h)$ in their commuting incoming tensor factors, the harmonic prescription reads
\begin{equation}
\begin{aligned}
&\eta_\sigma\big(:\!q(P_h+S_h)\!:\!_{\mathrm{probe},\sigma}\big)\\
&\quad\stackrel{\mathrm{formal}}{=}
\frac4\pi\sum_{r\geq0}\frac{(-1)^r}{2r+1}
\cos((2r+1)S_h)=q(S_h),
\end{aligned}
\label{eq:bosonic-formal-sign}
\end{equation}
where Fej\'er summation gives the rigorous meaning of the system-side limit in Eq.~\eqref{eq:bosonic-induced-limit}. The system field is not Wick ordered in this identity. In particular, the Gaussian smoothing of Eq.~\eqref{eq:bosonic-smoothed-sign} describes the ordinary, unordered sign readout, and is not an objection to the harmonic cancellation in Eq.~\eqref{eq:bosonic-ordered-weyl}.

For comparison, the ordinary sign readout without probe normal ordering behaves differently. In a normal product representation, Eq.~\eqref{eq:bosonic-noise} then gives, for $v_h>0$,
\begin{equation}
\begin{aligned}
\varepsilon_\sigma(q(\psi(h)))&=Q_{v_h}(\phi(f_h)),\\
Q_v(x)&=\sum_{n\ \mathrm{odd}>0}a_n e^{-n^2v/2}\cos(nx)\\
&=\int_{\mathbb R}\frac{e^{-y^2/(2v)}}{\sqrt{2\pi v}}\,q(x+y)\,dy.
\end{aligned}
\label{eq:bosonic-smoothed-sign}
\end{equation}
This series converges absolutely. The wrapped Gaussian assigns positive weight to both signs, so $\|Q_v\|_\infty<1$ for every $v>0$. Consequently the induced observable is a strict contraction, not an involution, and it is not $q(\phi(f_h))$. The latter identity holds in the exceptional noiseless case $v_h=0$; realizing that case for the required attainable modes is an additional dynamical and state condition.

This prescription subtracts only the variance of $\psi(g_h)$ after incoming pullback. Full vacuum Wick ordering of the combined outgoing field would also involve the system covariance and would not be the same prescription. For $v_h>0$ the inverse Gaussian factors grow as $e^{n^2v_h/2}$; with the $1/n$ square-wave coefficients they do not define a bounded periodic function in the infinite limit. In particular the Fourier coefficients of $B_{N,h}$ are unbounded as $N$ increases, precluding a uniformly bounded operator limit for a nonzero probe quadrature with spectrum $\mathbb R$. Thus the symbol $:\!q(\psi_{\mathrm{out}}(h))\!:\!_{\mathrm{probe},\sigma}$ may express formal harmonic cancellation, but no bounded readout with an exact sign--cosine output follows from it. Equation~\eqref{eq:bosonic-finite-subtraction} is valid at finite $N$; it neither preserves squares under the induced map nor supplies normalized binary probe outcomes.
In particular, Eq.~\eqref{eq:bosonic-induced-limit} is not an interchange of an infinite probe-series limit with $\varepsilon_\sigma$. An unbounded or generalized interpretation would require a specified domain and a justified expectation or summation rule beyond the bounded FV observable map used here.

\subsection{Attainable symplectic pairs, Pauli observables and Bell correlations}
To obtain exact system Pauli observables, suppose the real attainable range of the scalar output map contains $f,f'$ localized in Alice's region with
\begin{equation}
E_\phi(f,f')=(2\ell+1)\pi,\qquad \ell\in\mathbb Z.
\label{eq:bosonic-odd-phase}
\end{equation}
For all odd signed integers $m,n$, the Weyl commutation phase is $e^{-imnE_\phi(f,f')}=-1$. Hence all odd harmonics anticommute. Passing from the bounded Fej\'er approximants to their strong limits gives
\begin{equation}
\begin{gathered}
X_A=q(\phi(f)),\qquad Y_A=q(\phi(f')),\\
\{X_A,Y_A\}=0,\qquad X_A^2=Y_A^2=\mathbf1,\qquad Z_A=-iX_AY_A.
\end{gathered}
\end{equation}
These generate a Pauli algebra. Equation~\eqref{eq:bosonic-odd-phase} is the exact odd-Weyl-phase condition used for this construction in the convention~\eqref{eq:bosonic-weyl} \cite{FiniteWeyl2026}. It cannot be inferred from separate norm normalizations. Given an attainable pair $u,v$ with $\kappa=E_\phi(u,v)\neq0$, the real rescalings
\begin{equation}
f=\sqrt{\pi/|\kappa|}\,u,\qquad
f'=\operatorname{sgn}(\kappa)\sqrt{\pi/|\kappa|}\,v
\end{equation}
give $E_\phi(f,f')=\pi$ and preserve supports, provided their probe preimages remain admissible. If $\kappa=0$, rescaling cannot repair the pair. Nonzero symplectic pairing within each local attainable range is therefore a substantive assumption. Spacelike separation is imposed between the parties, not between the two directions within a party.

An analogous attainable pair in Bob's region gives $X_B,Y_B,Z_B$. Spacelike separated representatives have vanishing cross symplectic pairings, so the two Pauli algebras commute. The settings
\begin{equation}
\begin{gathered}
A=X_A,\qquad A'=Y_A,\\
B=(X_B+Y_B)/\sqrt2,\qquad B'=(X_B-Y_B)/\sqrt2
\end{gathered}
\end{equation}
are Hermitian involutions and yield
\begin{equation}
\mathcal B=\sqrt2(X_AX_B+Y_AY_B),\qquad
\|\mathcal B\|=2\sqrt2.
\end{equation}
Here the norm equality follows in the nonzero faithful representation of the two commuting matrix algebras. It is not an expectation in a specified vacuum or prepared state. Violation requires an admissible state with $|\omega(\mathcal B)|>2$ and a protocol realizing the joint readouts. Bob's rotated settings mix two noncommuting quadratures; four final settings each depending on a single quadrature in a centered quasifree state would instead admit the common Gaussian description discussed in \cite{FiniteWeyl2026}.

The finite Pauli-algebra argument of Appendix~\ref{app:pauli} also supplies a conditional singlet construction here. Set
\begin{equation}
\begin{gathered}
P_s=\frac14(\mathbf1-X_AX_B-Y_AY_B-Z_AZ_B),\\
U_0=\mathbf1,\qquad U_1=X_B,\qquad U_2=Y_B,\qquad U_3=Z_B.
\end{gathered}
\end{equation}
The four projections $U_\mu P_sU_\mu^*$ sum to $\mathbf1$. For any normalized vacuum vector $\Omega$, at least one $q_\mu=\|P_sU_\mu^*\Omega\|^2$ is positive, and
\begin{equation}
\Omega_s=P_sU_\mu^*\Omega/\sqrt{q_\mu}
\end{equation}
has singlet correlations $\langle X_AX_B\rangle_s=\langle Y_AY_B\rangle_s=\langle Z_AZ_B\rangle_s=-1$. With the displayed settings its CHSH expectation is $-2\sqrt2$; reversing both Bob settings gives $2\sqrt2$. This is an algebraic conditional preparation by a joint projection, not a demonstrated FV preparation protocol or a statement about the unprojected vacuum. Its applicability presupposes actual bounded system Pauli generators; a formal infinite normally ordered probe series does not by itself establish their implementation.

The scalar FV calculation thus supplies exact finite-harmonic first moments and a conditional route to the sign--cosine Pauli algebra. Positive probe noise prevents the ordinary sign readout from giving exact induced involutions; its damped odd harmonics can still anticommute when Eq.~\eqref{eq:bosonic-odd-phase} holds, but they do not form normalized Pauli generators. A complete measurement construction must establish the required local symplectic ranges, either noiseless transfer or controlled finite readout corrections, and joint statistics in a suitable state. Products and Bell correlations cannot be obtained by assuming multiplicativity of $\varepsilon_\sigma$.

\end{document}